\documentclass[preprint,12pt]{elsarticle}

\usepackage{amssymb}
\usepackage{lscape}

\journal{Computer Physics Communications}

\begin{document}

\begin{frontmatter}



\title{Comprehensive Machine Learning Model Comparison for Cherenkov and Scintillation Light Separation due to Particle Interactions}


\author[1,2]{Emrah Tiras\corref{cor1}}
\cortext[cor1]{corresponding authors}
\ead{etiras@fnal.gov}

\author[1]{Merve Tas\corref{cor1}}
\ead{mervetas@erciyes.edu.tr}

\affiliation[1]{organization={Department of Physics},
            addressline={Erciyes University}, 
            city={Kayseri},
            postcode={38030}, 
            country={Türkiye}}

\affiliation[2]{organization={Department of Physics and Astronomy},
            addressline={The University of Iowa}, 
            city={Iowa City},
            state={IA},
            postcode={52242},
            country={USA}}
            
\author[3]{Dilara Kizilkaya}

\affiliation[3]{organization={Department of Computer Engineering},
            addressline={Erciyes University}, 
            city={Kayseri},
            postcode={38039},
            country={Türkiye}}
            
\author[4]{Muhammet Anil Yagiz}

\affiliation[4]{organization={Department of Engineering and Natural Sciences},
            addressline={Kırıkkale University}, 
            city={Kırıkkale},
            postcode={71450},
            country={Türkiye}}
            
\author[5]{Mustafa Kandemir}

\affiliation[5]{organization={Department of Physics},
            addressline={Recep Tayyip Erdoğan University}, 
            city={Rize},
            postcode={53100},
            country={Türkiye}}

\begin{abstract}
The demand for novel detector mediums such as Water-based Liquid Scintillator (WbLS) has increased over the last few decades due to their capability for both low energy particle interactions and higher light yield. Recently, the usage of machine learning (ML) methods in high-energy physics has also been increasing. The ML and AI methods are used in many physics projects in the field since they provide effective and sensitive results. In this study, we aimed to develop a comprehensive analysis of water Cherenkov detectors and perform physics analyses to efficiently separate Cherenkov and scintillation photons with ML algorithms using the data from the WbLS detector environment. The main goal of this study was to produce more precise solutions to physics problems, such as signal classification, by applying ML techniques to the simulation and experimental data. Here, we trained more than 20 ML models, and our results revealed that three machine learning models, XGBoost, Light GBM, and Random Forest models, and their ensemble model gave us more than 95\% accuracy for separating Cherenkov and scintillation photons with balanced and unbalanced datasets. This is a significant increase in efficiency as compared with the results of the classical method by applying simple time cuts.

\end{abstract}




\end{frontmatter}


\section{Introduction}
\label{Introduction}

Since neutrinos interact only weakly with matter and matter particles, we need massive and extremely sensitive neutrino detectors to detect many neutrino interactions. For this purpose, water- and liquid-based massive optical detectors have been used successfully in neutrino physics \cite{Becker-Szendy:1995qpb, KamLAND:2002uet, Seo:2019shs}. Over the last few decades, the demand for novel detector mediums such as Water-based Liquid Scintillator (WbLS), which consists of a mixture of a small volume of liquid scintillator (LS) with a large volume of water, has been increasing due to their capability for both low energy particle interactions and higher light yield. The high percentage of water in the medium provides a low manufacturing cost and easier experimental handling of the detector.

These hybrid detectors simultaneously detect Cherenkov and scintillation light to enhance reconstruction and particle identification (PID) capabilities \cite{aberle2014measuring, elagin2017separating}. They offer a high light yield and low energy threshold and increase the direction reconstruction capacity \cite{Kaptanoglu:2021prv}. Cherenkov radiation helps determine particle direction, track reconstruction, and particle identification via the Cherenkov ring's topology. Adding a small portion of the scintillator enhances energy resolution and facilitates the detection of particles below the Cherenkov threshold.

The separation of Cherenkov and scintillation light is crucial in high-energy physics \cite{article}. This separation is significant for originating the coming light's physical process, reconstruction, enhancing the identification of the coming particles, and differentiating signal from the noise. In recent years, many different approaches have been developed for this separation. These approaches are based on spectral photon sorting \cite{kaptanoglu2019cherenkov, kaptanoglu2020spectral}, time separation using fast photodetectors \cite{aberle2014measuring, elagin2017separating, gruszko2019detecting} and innovation of new target materials, such as slow scintillators \cite{dunger2022slow, guo2019slow} and WbLS \cite{caravaca2020characterization, yeh2011new}. WbLS is an innovative target material that enables the simultaneous detection of Cherenkov radiation and scintillation. 

The Accelerator Neutrino Neutron Interaction Experiment (ANNIE) \cite{back2017accelerator, back2020measurement} collaboration deployed both WbLS and Large Area Picosecond Photodetectors (LAPPDs) to improve vertex reconstruction and neutron detection efficiency \cite{ANNIE:2023yny}. Kaptanoglu et al. \cite{Kaptanoglu:2021prv} demonstrated the combination of WbLS and LAPPDs to detect Cherenkov light from a hybrid medium using a low-energy source and a fast-timing response. V. Fischer and E. Tiras \cite{Fischer:2020htg} investigated a neutrino detector to create a testbed for WbLS and  LAPPDs. The proposed method detects Cherenkov light with high purity. Caravaca et al. \cite{caravaca2020characterization} measured the scintillation light yield and time profile for three WbLS mixtures at different LS concentrations in water using the Monte Carlo (MC) model. They improved performance for separating Cherenkov and scintillation signals for cosmic muons in a small-scale target (centimeters). Another work by Caravaca et al. \cite{caravaca2017cherenkov} proposed Cherenkov and scintillation light separation in organic liquid scintillators. They performed time and charge-based separation of Cherenkov and scintillation light in organic liquid scintillators. 

Separation methods such as classical ones, like applying simple time cuts, are commonly used \cite{Kaptanoglu:2021prv}, but they may result in data loss and do not provide sensitive results. In addition to the classical methods, machine learning (ML) techniques have been vital in high-energy physics \cite{Jamieson:2022scv, Bourilkov:2019yoi} and especially particle physics for almost two decades \cite{brun1997root}. ML methods in particle physics experiments are highly efficient and expand particle physics research \cite{Psihas:2020pby}. Gavrikov et al. \cite{gavrikov2022energy} proposed ML techniques, such as Boosted Decision Trees (BDT) and Fully Connected Deep Neural Networks (FCDNN), for precise energy reconstruction in the energy range of 0–10 MeV. Madden et al. \cite{madden2018temporal} proposed an artificial intelligence (AI) model for temporal separation of Cherenkov radiation and scintillation. The Convolutional Neural Network (CNN) was trained to predict Cherenkov radiation in the temporal response of a LINAC irradiated scintillator-fiber optic dosimeter. They achieved satisfactory performance matching the background subtraction method. Lai et al. \cite{Lai:2020byl} proposed an ML framework for parton showers using Generative Adversarial Networks (GANs). Abratenko et al. \cite{MicroBooNE:2020hho} proposed a multiple particle identification (MPID) network for the MicroBooNE experiment. This network enhances the performance of MicroBooNE's already in-use single particle identification (PID) CNN network. The network's scores were sufficient in both simulation and beam data. 

In the present study, we investigate the classification of Cherenkov and scintillation light using classical methods and ML techniques. In addition to this analysis, the performance of several ML models is compared. This study is significant for future classification problems in neutrino physics.  

\section{Simulation Studies}
\label{simulation}
In the scope of this study, we developed a generic neutrino detector using the Geant4 Monte Carlo simulation package \cite{Agostinelli}. We also utilized G4DCP \cite{g4dcp} to achieve a more flexible detector setup and OPSim \cite{opsim} to simplify the implementation of optical photons. We consolidated various simulation settings into a macro file, enabling a unified interface to manage the entire simulation easily.

\begin{figure}[h!]
    \centering
    \includegraphics[width=0.7\textwidth]
{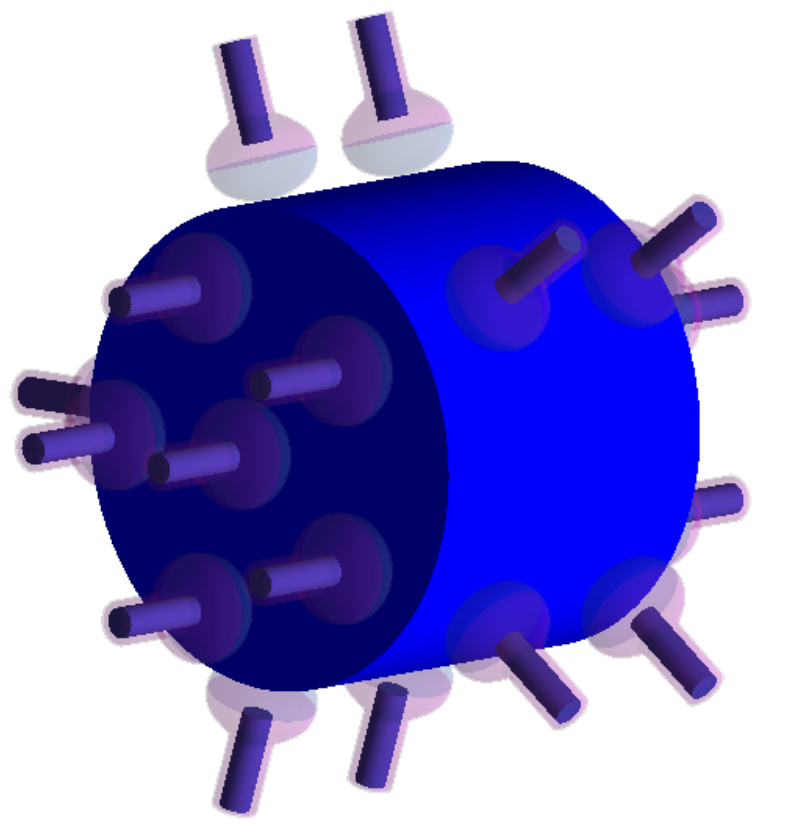}
    \caption{Simulated detector model obtained from Geant4.}
    \label{SimFig1}
\end{figure}

Using the developed interface, we designed a custom detector geometry comprising 21 Photomultiplier Tubes (PMTs) surrounding a cylindrical volume with a diameter of 866 mm and a height of 654 mm. The detector volume was filled with WbLS, a novel low-energy threshold detection medium that perfectly bridges the scintillator and water. A sketch of the detector model is shown in Figure \ref{SimFig1}.

We initiated the simulation by firing electrons with an energy of 1 MeV from the central point of the detector volume, directed towards the negative z-axis. As simulation output, we recorded various physics quantities on a per-event basis for subsequent analysis, including the number of emitted Cherenkov and scintillation photons and the type, energy, and arrival time of photons detected by each PMT and the PMT coordinates. 

\begin{figure}[h!]
    \centering
    \includegraphics[width=1.0\textwidth]
{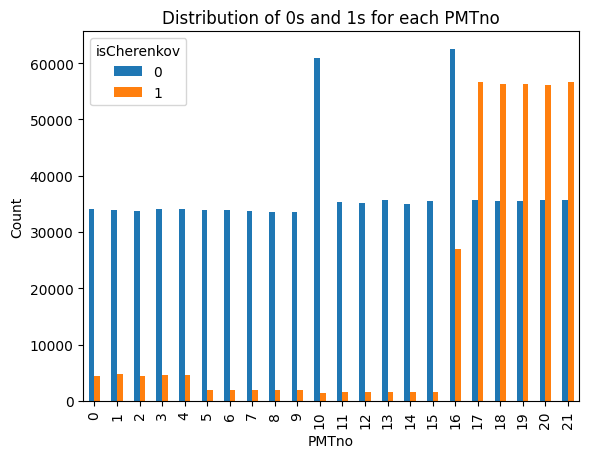}
    \caption{The distribution of Cherenkov and scintillation photons detected by PMTs.}
    \label{SimFig2}
\end{figure}

Fig. \ref{SimFig2} shows the distribution of the number of Cherenkov and scintillation photons detected by each PMT. As expected, PMTs positioned at the bottom, aligned with the direction of electron ejection, detected a higher number of Cherenkov photons than the other PMTs.

\begin{figure}[h!]
    \centering
    \includegraphics[width=1.0\textwidth]
{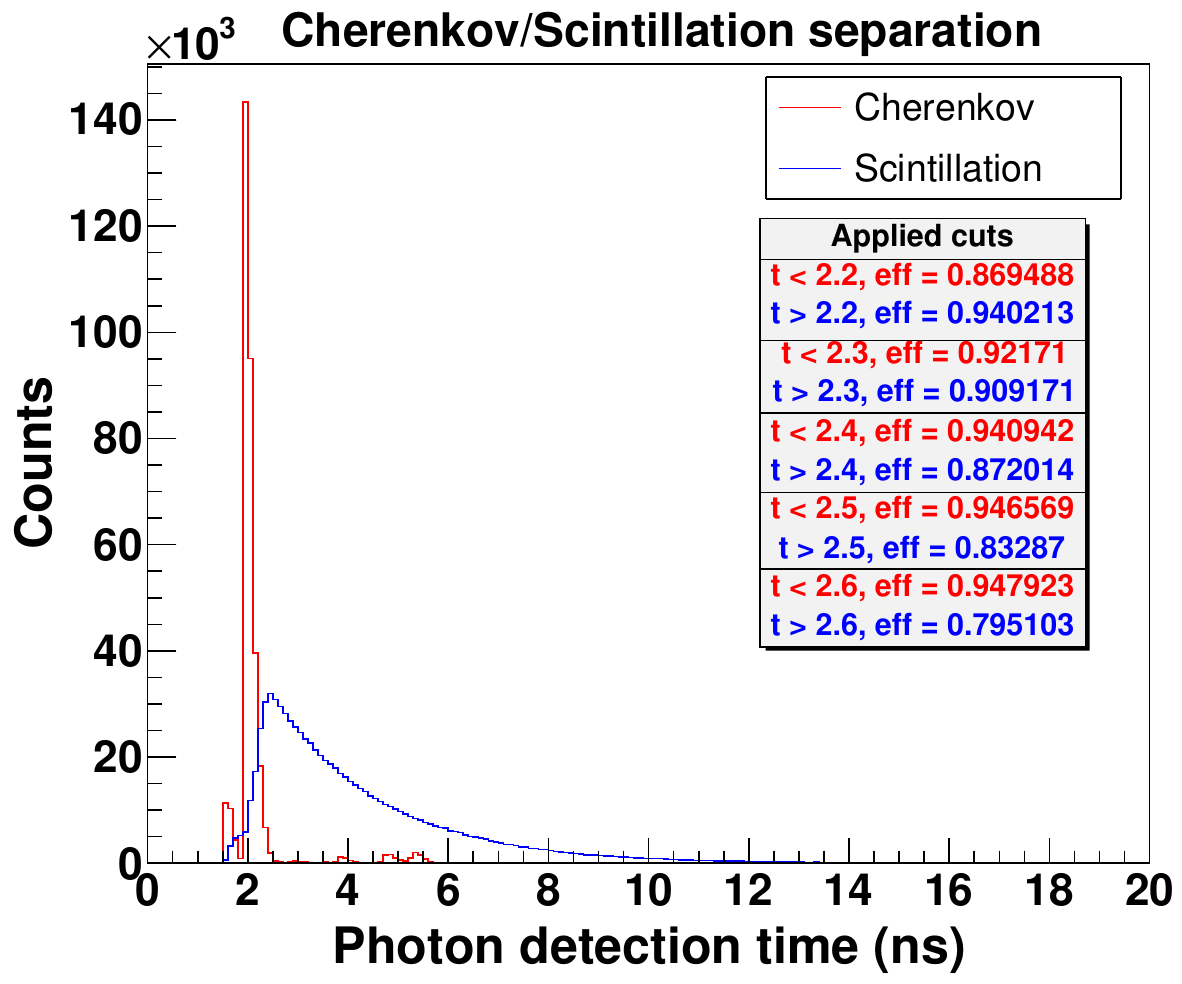}
    \caption{Estimated selection efficiency of Cherenkov and scintillation lights based on the applied cuts.}
    \label{SimFig3}
\end{figure}

Before using ML algorithms, we applied various time cuts to the simulation data to estimate the selection efficiency of Cherenkov and scintillation photons using the classical method. Fig. \ref{SimFig3} displays the efficiency results based on the applied cuts.

\section{Machine Learning Methods and Classification}
\label{MLmethods}
\subsection{Dataset Preparation}
\label{Dataset}
    The Geant4 simulation platform generates the output data through various means, with the most common formats being ROOT and CSV files. We pre-processed the data generated from the Geant4 simulation platform to prepare it for input into the ML-based classification system. After pre-processing, the obtained CSV file is appropriate for use as input data.

    Unedited data consisted of five different features: event numbers, PMT IDs providing PMT coordinates, photon arrival time, photon hit energy, and radiation type of corresponding photons. Event number is not included in the final dataset because electrons come from one direction, meaning that regardless of the event number,  the distribution of Cherenkov and scintillation light detected by each PMT is consistent across every event.

    We prepared two different datasets, balanced and unbalanced. The balanced dataset has 700k data (number of photons in all events); half is scintillation data, and the other is Cherenkov data. The unbalanced dataset has over 1M of approximately 350k Cherenkov data and 810k scintillation data.
 
\subsection{Model Selection}
\label{Model}

Recently, ML techniques have been widely used for classification problems. In this context, we utilized the most common ML models in the literature to directly classify Cherenkov and scintillation radiation. First, we worked on many ML-based classification models. Then, we chose the 13 most effective ML classifier models based on their classification accuracy score for our Cherenkov/scintillation dataset. All these models are mentioned in  Section \ref{Results}. Three models, Light GBM, XGBoost, and Random Forest, have performed better than the other models in our case.  

The Random Forest algorithm, proposed by L. Breiman \cite{Breiman:2001hzm}, has been successful in classification and regression methods. This approach, which integrates multiple randomized decision trees and combines their predictions through averaging, has demonstrated outstanding performance in scenarios where the number of variables greatly exceeds the number of observations. Gradient boosting, proposed by Chen Tianqi \cite{Chen:2016btl}, is based on training weak learners independently of each other and developing the next learner by learning from the mistakes of these learners. Gradient Boosting Machine (GBM) aims to minimize the error between the predicted values of each learner and the actual values. XGBoost (eXtreme Gradient Boosting) is an enhanced version of GBM. XGBoost provides a faster solution by utilizing parallel computations on large datasets compared to GBM. Additionally, XGBoost achieves high-accuracy predictions in less computational time than other algorithms. Microsoft also developed Light GBM to obtain high performance on large datasets. Light GBM \cite{ke2017lightgbm} uses a histogram-based learning technique to provide faster data processing with less memory. 

\begin{figure}[h]
\centering
\includegraphics[width=0.8\textwidth]{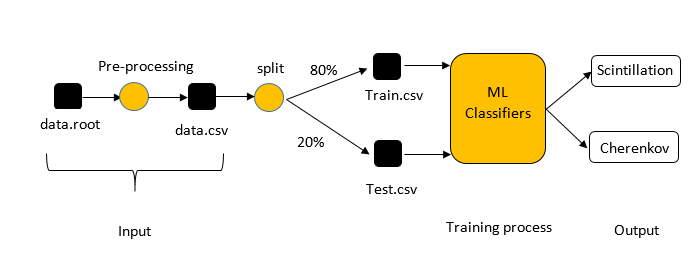} 
\caption{The graphical representation of the training process and the analysis.}
\label{MLfig1}
\end{figure}

\begin{figure}[h!]
    \centering
    \includegraphics[width=1.0\textwidth]
{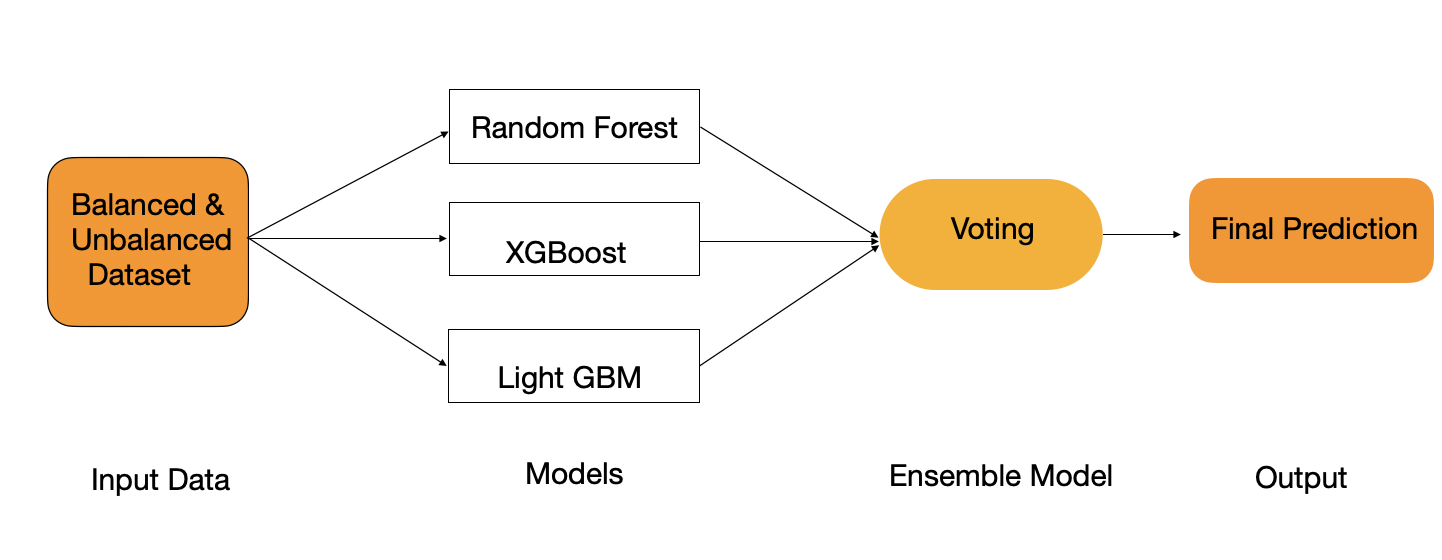}
    \caption{The graphical structure of the ensemble model.}
    \label{MLfig2}
\end{figure}

Figure~\ref{MLfig1} shows the graphical representation of the training process and the analysis. Moreover, the grid-search hyper-parameter tuning technique was performed on the three best models to improve their accuracy. Finally, we obtained the ensemble model, which combines the three most successful models with the voting classifier method. Hyper-parameter tuning was also performed with the ensemble model and the best parameter set was used for this task. Figure~\ref{MLfig2} demonstrates the structure of the ensemble model.

We also trained ML-based models for two separate balanced and unbalanced datasets. All the input parameters selected for balanced and unbalanced datasets are listed in Tables \ref{MLtable1} and \ref{MLtable2} in Section~\ref{Results}. 

\subsection{Evaluation Metrics}
\label{Evaluation}

Precision (P), Recall (R), F1 score, and Accuracy metrics are used to evaluate the system performance. When the Cherenkov radiation is predicted correctly according to ground truth, it is regarded as True Positive (TP). In the False Positive (FP) case, the system predicts a Cherenkov radiation that does not match the ground truth. True Negative (TN) occurs when the model predicts scintillation radiation correctly according to ground truth. Prediction of a scintillation radiation in a test set that does not match the true label is regarded as False Negative (FN).

\begin{equation}
   Precision(P) =\frac{TP}{TP+FP}
\end{equation}
\begin{equation}
   Recall (R) =\frac{TP}{TP+FN}
\end{equation}
\begin{equation}
   F1\,Score = \frac{2*P*R}{P+R}
\end{equation}
\begin{equation}
   Accuracy = \frac{TP+TN}{TP+FP+TN+FN}
\end{equation}

Another evaluation metric is the logarithmic or cross-entropy loss (log-loss) function, which represents the reliability of the models. It measures the performance of the classification model by quantifying the accuracy of the probabilities it predicts. Equation 5 shows the log-loss calculation. 

\begin{equation}
   Log Loss =-\frac{1}{N}\sum_{i=1}^{N}[y_i.\log(p_i)+(1-y_i).\log(1-p_i)]
\end{equation}

in which N is the number of instances, $y_{i}$ is the true label ith instance, and $p_{i}$ is the predicted probability of the $i^{th}$ instance belonging to a positive class.

\section{Machine Learning Model Comparisons and Results}
\label{Results}

We present our classification results on two separate datasets, balanced and unbalanced. We also performed input parameters such as time, energy, PMT coordinates, and their combinations. For the classification of Cherenkov and scintillation photons, we trained more than 20 state-of-the-art ML classification models according to their accuracy scores. These models are classifiers based on tree models (Decision Tree, Extra Tree), linear models (Logistic Regression, Ridge Classifier), ensemble models (Adaboost, Random Forest, Extra Trees), boosting models (XGBoost, Light GBM), and other models such as KNeighbors and Support Vector Machine. 

\begin{figure}[h!]
    \centering
    \includegraphics[width=1.0\textwidth]
{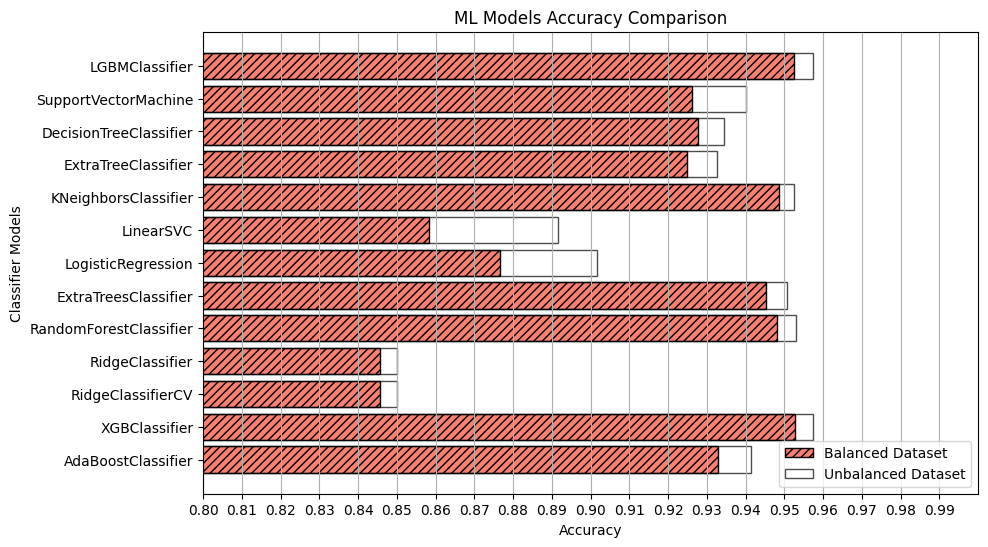}
    \caption{Comparison of the accuracy of all models for balanced and unbalanced datasets with accuracy above 80\%.}
    \label{MLfig3}
\end{figure}

Both balanced and unbalanced datasets were utilized in the training and testing process with an 80\% train and a 20\% test data split. Figure \ref{MLfig3} shows all the models with 80\% or higher accuracy. Five of these models (Random Forest, XGBoost, Light GBM, KNeighbors, and ExtraTrees) with both balanced and unbalanced datasets provided more than 94\% accuracy, and three of them (Random Forest, XGBoost and Light GBM) with the highest accuracy were selected for further optimization.  

We selected these three models with the highest accuracy and performed a grid-search hyper-parameter optimization technique. Then, we proposed an ensemble method with these three classification models to enhance model performance.

\begin{figure}[h!]
    \centering
    \includegraphics[width=1.0\textwidth]
{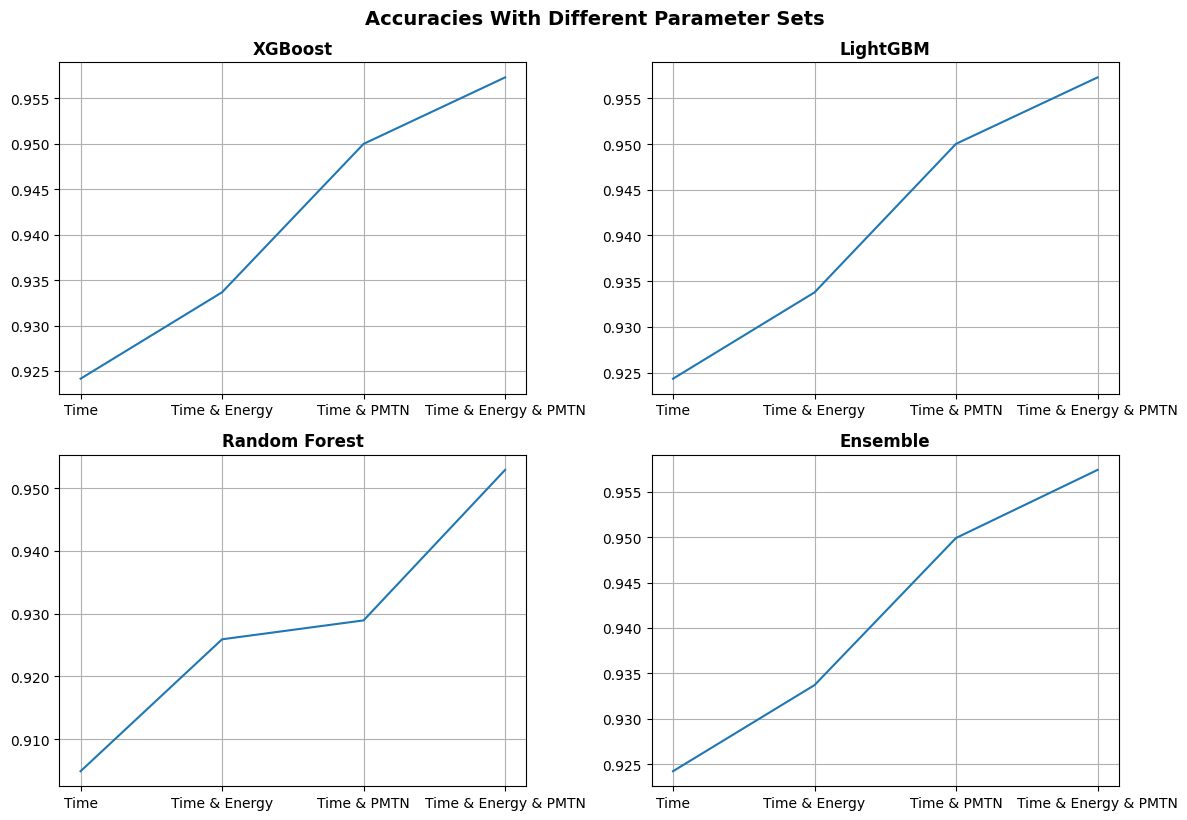}
    \caption{Comparison of the accuracy of the three best models and their ensemble for different parameters.}
    \label{MLfig4}
\end{figure}

Figure~\ref{MLfig4} shows the three best models and their ensemble model accuracy for four different input parameters and their combinations. The results showed that the accuracy of the models varies based on input parameters. All models achieve their highest accuracy when time, energy, and PMT coordinates are available together as input parameters. The models also achieved accuracy scores higher than 0.90\% when only the time is used and 0.93\% when time and energy are used together as input parameters. Thus, even if some parameters are missing in the experimental data, the ML techniques would provide higher classification accuracy than the existing parameters for particle physics experiments. 

Compared to classical methods, our results for separating Cherenkov and scintillation photons are quite successful. We reached approximately 0.96\% of classification accuracy. All model accuracy results corresponding to different input instances are summarized in Table~\ref{MLtable1}. Table~\ref{MLtable1} gives Cherenkov and scintillation photons classification performances obtained from different ML classification models with various parameter sets while using balanced and unbalanced datasets. The best three models and their ensemble model outputs are shown in Table~\ref{MLtable2}. The classification performances in terms of accuracy on an unbalanced dataset yielded better
results when compared to a balanced dataset case.

\begin{figure}[h!]
    \centering
    \includegraphics[width=1.0\textwidth]
{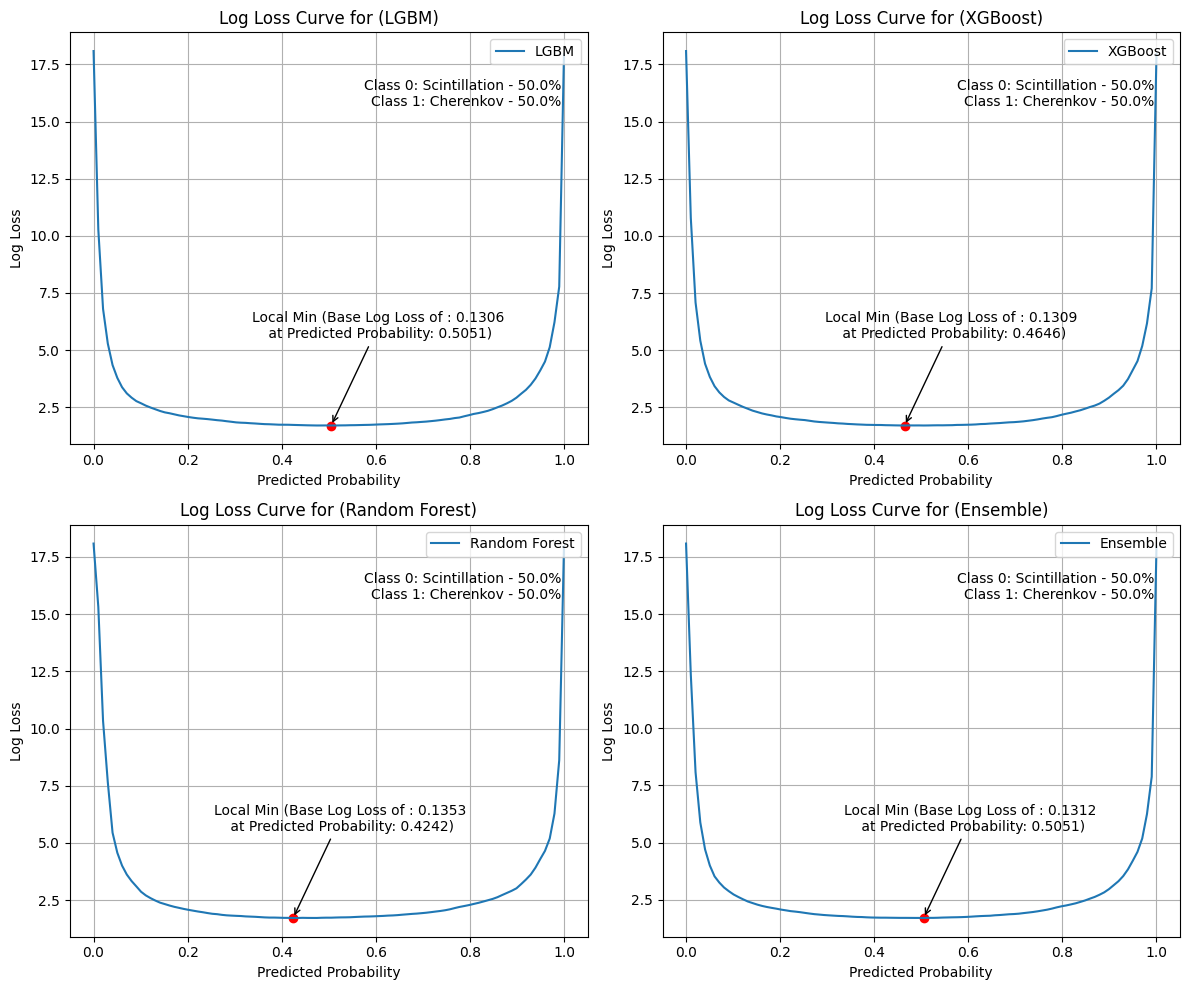}
    \caption{Analysis of the log-loss curve for ML models (LGBM, XGBoost, Random Forest, and Ensemble) on a balanced dataset.}
    \label{MLfig5}
\end{figure}

\begin{figure}[h!]
    \centering
    \includegraphics[width=1.0\textwidth]
{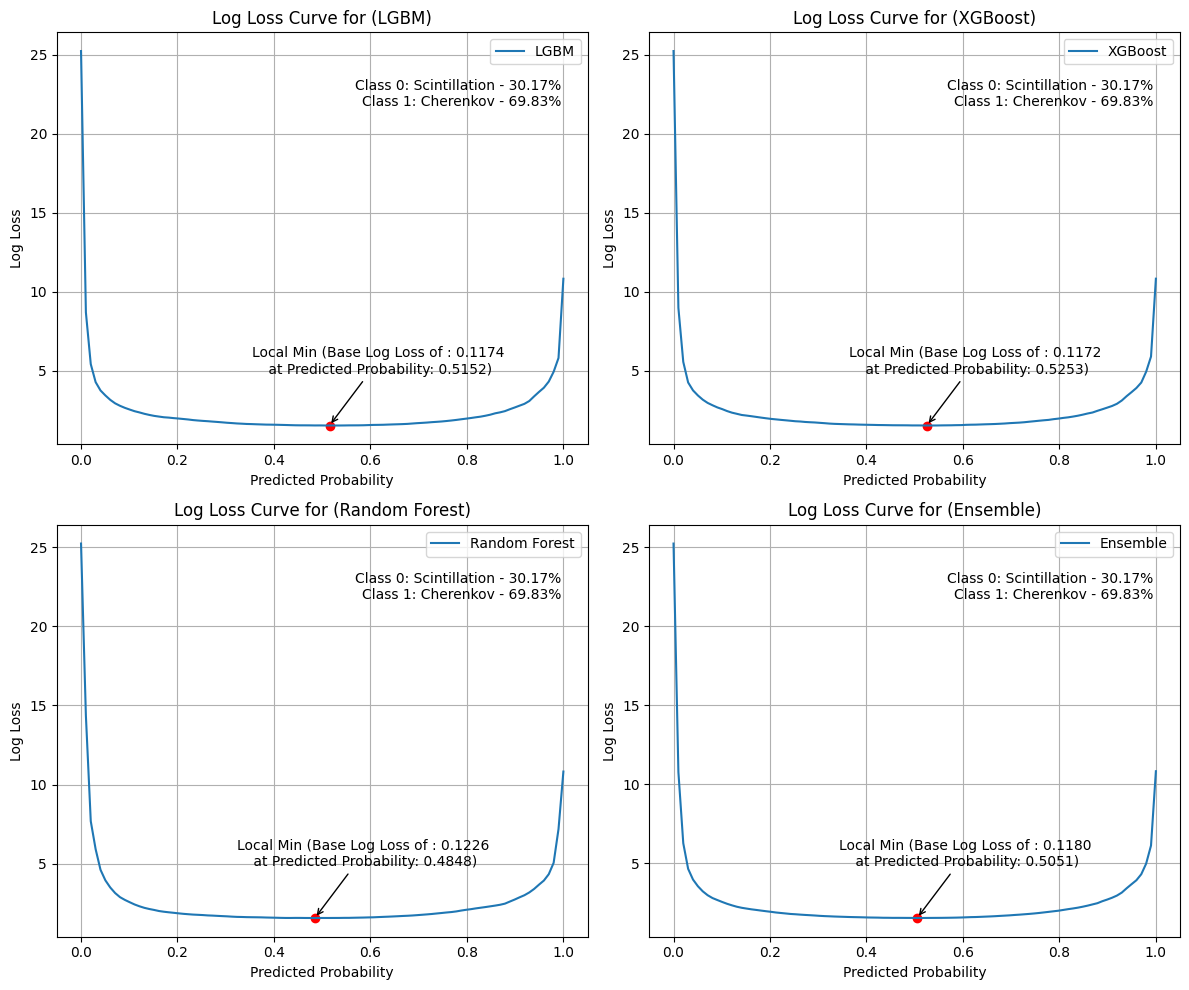}
    \caption{Analysis of the log-loss curve for the ML models (LGBM, XGBoost, Random Forest, and Ensemble) on an unbalanced dataset.}
    \label{MLfig6}
\end{figure}

Figures \ref{MLfig5} and \ref{MLfig6} depict the analysis of each model's log-loss score on balanced and unbalanced datasets, respectively. The log-loss score measures how closely the predicted probability matches the actual value. Both classes will have similar weights in log-loss calculations in the balanced dataset with nearly equal class distribution. In an unbalanced dataset, where there is a difference in the number of observed samples between classes, the minority class has a higher impact on log-loss calculations.  The local minimums on the balanced dataset for Light GBM, XGBoost, Random Forest, and Ensemble are 0.1306, 0.1309, 0.1353, and 0.1312, and their corresponding predicted probabilities are 0.5051, 0.4646, 0.4242 and 0.5051, respectively. The local minimums on the unbalanced dataset for Light GBM, XGBoost, Random Forest, and Ensemble are 0.1174, 0.1172, 0.1226, and 0.1180, and their corresponding predicted probabilities are 0.5152, 0.5253, 0.4848 and 0.5051, respectively. A higher log-loss score means low accuracy while a low log-loss score indicates higher accuracy. 

\begin{landscape}
\vspace*{2cm}
\begin{table}[h]
\caption{Accuracy values corresponding to all ML classifiers with different input parameters.} 
\centering
\begin{tabular}{ |c|c|c|c|c|c|c|c|c|c| } 
    \hline
    ML Classifier & \multicolumn{2}{c|}{Time} & \multicolumn{2}{c|}{Time-Energy} & \multicolumn{2}{c|}{Time-PMTN} & \multicolumn{2}{c|}{Time-Energy-PMTN} \\ 
    \hline
    & Balanced & Unbalanced & Balanced & Unbalanced & Balanced & Unbalanced & Balanced & Unbalanced \\ 
    \hline
    AdaBoost & 0.9152 & 0.9236 & 0.9222 & 0.9272 & 0.9168 & 0.9396 & 0.9329 & 0.9414 \\ 
    \hline
    Logistic Regression & 0.8607 & 0.9100 & 0.8711 & 0.8810 & 0.8535 & 0.9016 & 0.8766 & 0.9016 \\
    \hline
    K-Nearest Neighbors & 0.9069 & 0.9150 & 0.9211 & 0.9261 & 0.9390 & 0.9445 & 0.9485 & 0.9526 \\
    \hline
    Ridge ClassifierCV & 0.7791 & 0.8196 & 0.8008 & 0.8196 & 0.8051 & 0.8981 & 0.8457 & 0.8501 \\
    \hline
    Ridge Classifier & 0.7791 & 0.8196 & 0.8008 & 0.8196 & 0.8051 & 0.8981 & 0.8457 & 0.8501 \\
    \hline
    Extra Trees & 0.8875 & 0.9040 & 0.9176 & 0.9246 & 0.9183 & 0.9269 & 0.9454 & 0.9507 \\
    \hline
    Linear SVC & 0.8350 & 0.9003 & 0.8640 & 0.8667 & 0.8872 & 0.8700 & 0.8583 & 0.8917 \\
    \hline
    Support Vector Machine & 0.9003 & 0.9238 & 0.9206 & 0.9238 & 0.9243 & 0.9350 & 0.9263 & 0.9400 \\
    \hline
    Extra Tree & 0.8871 & 0.9037 & 0.8900 & 0.9004 & 0.9173 & 0.9257 & 0.9249 & 0.9326 \\
    \hline
    Decision Tree & 0.8875 & 0.9039 & 0.8910 & 0.9008 & 0.9163 & 0.9248 & 0.9277 & 0.9345 \\
    \hline
\end{tabular}
\label{MLtable1}
\end{table}


\begin{table}[h]
\caption{Accuracy values for three best ML models and their ensemble model using different input parameters.} 
\centering

\begin{tabular}{ |c|c|c|c|c|c|c|c|c|c| } 
    \hline
    ML Classifier & \multicolumn{2}{c|}{Time} & \multicolumn{2}{c|}{Time-Energy} & \multicolumn{2}{c|}{Time-PMTN} & \multicolumn{2}{c|}{Time-Energy-PMTN} \\ 
    \hline
    & Balanced & Unbalanced & Balanced & Unbalanced & Balanced & Unbalanced & Balanced & Unbalanced \\ 
    \hline
    Random Forest & 0.8909 & 0.9049& 0.9211 & 0.9265 & 0.9203 & 0.9289 & 0.9510 & 0.9530 \\ 
    \hline
    XGBoost & 0.9170 & 0.9241 & 0.9285 & 0.9336 & 0.9448 & 0.9501 & 0.9526 & 0.9573 \\
    \hline
    Light GBM & 0.9170 & 0.9242 & 0.9286 & 0.9336 & 0.9449 & 0.9502 & 0.9528 & 0.9574 \\
    \hline
    Ensemble & 0.9170 & 0.9243 & 0.9287 & 0.9337 & 0.9449 & 0.9502 & 0.9529 & 0.9575 \\
    \hline
\end{tabular}
\label{MLtable2}
\end{table}

\end{landscape}

\section{Conclusion and Discussion}
\label{Conclusion}

This study comprehensively evaluated ML models for classifying Cherenkov and scintillation photons due to neutrino interactions. Several ML models were compared in balanced and unbalanced datasets, and their accuracies were calculated. All the related parameters and their combinations such as time, energy, and PMT coordinates in the detector were carefully studied. Our results revealed that ML-based classification methods provide better separation results for Cherenkov and scintillation photons when compared with the classical methods. The best results were received from the Random Forest, XGBoost, and Light GBM models, as well as the ensemble of these three models. The accuracy results with the unbalanced dataset are 95.30\%, 95.73\%, 95.74\%, and 95.75\%, with the balanced dataset being 95.10\%, 95.26\%, 95.28\%, and 95.29\%, respectively. Compared to the result of the classical method, an almost 6\% increase in accuracy using ML models is crucial for particle interactions, especially neutrino studies. 

In the scope of this study, the simulation was conducted by using directional beam interaction with the simulated detector. Directional information is useful primarily for accelerator and fixed-location experiments. However, as a separate study, we plan to expand the ML-based classification study with isotropic beam interaction. Furthermore, a new study with CNN will be performed to process the images of the reconstructed particle interactions.

\section{Acknowledgement}
\label{Ack}
This work was supported by Scientific Research Projects (BAP) of Erciyes University, Türkiye, under the grant contracts of FBAÜ-2023-12325, FBA-2022-12207, FBG-2022-11499, and FDS-2021-11525. Dr. Emrah Tiras is thankful for the support of the Turkish Academy of Sciences (TUBA) under the Outstanding Young Scientists Awards Program (GEBIP) grant. We thank the Office of the Dean for Research for providing the Lab’s infrastructure at the ARGEPARK building of Erciyes University and the Proofreading \& Editing Office for proofreading this manuscript.





\clearpage

\bibliographystyle{elsarticle-num}

\bibliography{sample}


\end{document}